\documentclass[fleqn,10pt]{wlscirep}
\usepackage[utf8]{inputenc}
\usepackage[T1]{fontenc}

\title{Spectra-Scope: A toolkit for automated and interpretable characterization of material properties from spectral data}

\author[1]{Amalya C. Johnson}
\author[1]{Chris Fajardo}
\author[1]{Leena Sansguiri}
\author[1]{Weike Ye}
\author[1,*]{Steven B. Torrisi}
\affil[1]{Energy \& Materials Division, Toyota Research Institute, Los Altos, CA 94022}
\affil[*]{steven.torrisi@tri.global}

\begin{abstract}
Spectroscopy is a central pillar of materials characterization, providing useful information on properties like structure, composition, or excited state dynamics of a system. However, many spectroscopic techniques present challenges in the development of interpretable, performant, and reliable supervised learning models due to the wide range of possible nonlinear correlations that can exist between the signal and the response variable (target) of interest. Here, we present Spectra-Scope, an open-source AutoML framework for automatic characterization of material properties from spectroscopy data using interpretable machine learning (ML) models. The software is implemented in Python and a no-code web application. It comprises tools for data preprocessing, nonlinear feature extraction, machine learning model training, and feature selection. Users can easily train different types of simple, interpretable ML models on a set of feature transformations quickly and with modest computational resources. In this work, we outline the methods of Spectra-Scope and its effectiveness across diverse datasets, with applications to materials and agricultural spectroscopy data. We show that Spectra-Scope can reproduce performance of comparable models in the literature, and highlight how our emphasis on interpretability can be used to rationalize the behavior of individual models and understand the physical processes behind spectral features. 

\end{abstract}
\begin{document}

\flushbottom
\maketitle
%
%
\thispagestyle{empty}


\section*{Introduction}

Spectroscopy is a powerful method for scientific analysis that probes the interaction of matter with electromagnetic radiation. It allows for the investigation of electronic, structural, and dynamic properties of physical systems, and is widely employed across disciplines such as materials science, chemistry, physics, and the life sciences. Autonomous and high-throughput laboratories are increasingly popular for exploring the synthesis process space for materials \cite{szymanski_autonomous_2023, gregoire_combinatorial_2023, baird_building_2023, ebrahimi_advancing_2024, chao_cuttingedge}. These high-throughput experimental schemes necessitate companion analysis pipelines for characterization of the complex datasets they produce.

Automatic spectroscopic analysis tools have thus become a key component of the high-throughput experimentation pipeline. This analysis can span from simple peak fitting \cite{takeuchi_datamanagement} to advanced machine learning (ML) \cite{ogunlade_2024, solis_mltblg, SHEREMETYEVA2020455} depending on the experimental technique and ultimate goal. Spectroscopic models can enable real-time feedback that supports decision-making during experiments \cite{liang_realtime}, enhances the detection of subtle features and patterns in data \cite{zhang_identifying_2020}, and alleviates the burden of repetitive manual analysis \cite{joy_simple_2022}. Thus, designing robust, interpretable, and computationally efficient models for spectroscopic data analysis is a ubiquitous and important challenge in developing high-throughput experimentation workflows.

Automated machine learning (AutoML) software pipelines streamline and automate common ML tasks like data preparation, feature engineering, model training, and hyperparameter tuning. While AutoML is a well-established technique in machine learning communities, few tools exist to apply and optimize such pipelines for materials-specific data. Previous works have applied AutoML techniques to specific materials systems or experimental pipelines, but lack generalizability to other experimental techniques or materials datasets \cite{tsamardinos_automated_2020, ji_research_2022}. More general-purpose frameworks have been established for manufacturing or time-series data \cite{sun_smart_2021, sun_alven_2020, tsfresh}. To the best of our knowledge, the literature lacks a versatile toolkit that seamlessly integrates existing methods for automated feature generation, model training, feature selection, and inference for spectroscopy data. Exposing these functionalities in a common package and interface might help accelerate the model development and data interrogation process for both experts and novice users.

\begin{figure}
    \centering
    \includegraphics[width=1\linewidth]{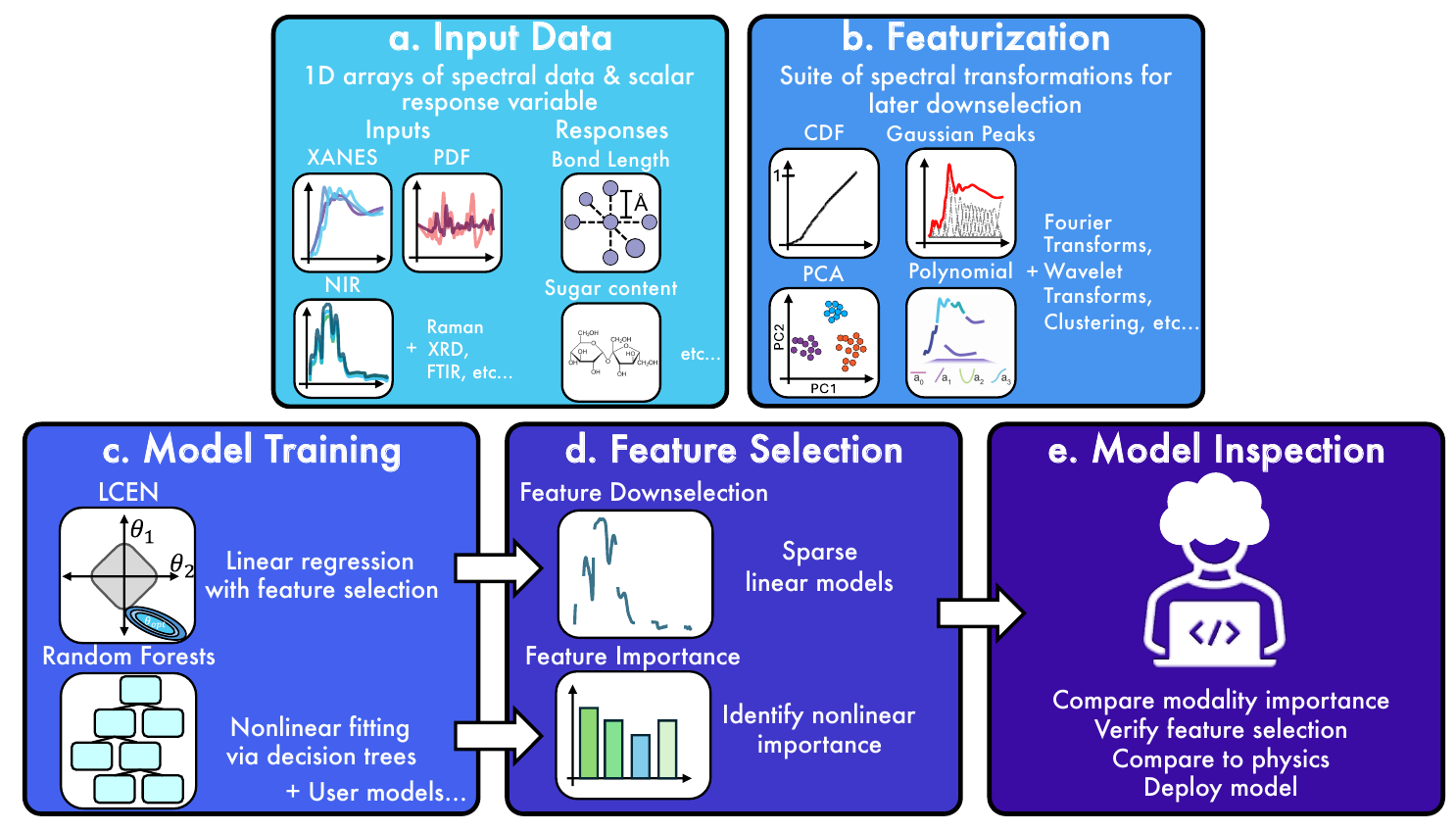}
    \caption{\textbf{Outline of this paper and the Spectra-Scope pipeline.} (a) Input data can come from any experimental or simulated 1-D array data source for inference on a scalar response variable. (b) Available featurizations of spectral data include the cumulative distribution function, Gaussian peak fitting, principal component analysis, polynomial peak fitting, and others as outlined in the methods. (c) Transformed spectra are used to train a machine learning algorithm. This paper focuses on the LCEN algorithm and random forests, but Spectra-Scope can be incorporated with user-built models as well. (d) Model training of LCEN and random forests includes feature selection either in the form of LCEN feature selection or random forest feature importances. (e) The algorithms are used to predict the input response variables. Feature selection helps with model interpretability and investigating modality importance.}
    \label{fig:1}
\end{figure}

This work describes Spectra-Scope, an open-source AutoML framework with a Python and web app implementation that automates simple supervised learning model development from spectral data. It comprises three main capabilities: (i) A library of spectral featurizers that can help transform raw spectra into features that better correlate with target properties, (ii) Model training using random forests \cite{breiman_random_2001} and regularized linear regression, both of which offer handles for interpretability, and (iii) Demonstrated support for different simultaneous modalities of data. In Spectra-Scope, the feature engineering methods chosen are specifically designed for supervised learning tasks on spectroscopy data, with interpretability of physical phenomena and parsimony of generated features as primary design choices. This differentiates Spectra-Scope as an AutoML framework specifically for spectroscopy. The capabilities demonstrated here only require modest computational resources. The overall workflow is described in \autoref{fig:1}. Full details are provided in the Methods section.

\section{Methods}
In this section, we will walk through three main features of the work: (i) the featurization methods we use, (ii) the models that we implement, and (iii) an overview of the application.

\subsection{Featurization}
We present a bundled implementation of ``featurizers'' to transform input spectra, extracting features that may better correlate with a property of interest than the raw spectra. This compiles techniques seen across the literature in one place. This allows practitioners to screen over featurizations with no \emph{a priori} knowledge of the expected relationships between spectral signal and response variables. Examples include transformations such as the cumulative distribution function \cite{chen_robust_2024} and multi-scale polynomial fitting \cite{torrisi_random_2020}.

We categorize our transformations and features into three conceptual sets: 1. ``Local'' features (that capture information in finite neighborhoods along the spectrum, e.g., polynomial fitting \cite{torrisi_random_2020}), 2. ``Nonlocal'' features (which incorporate information from the entire spectrum, e.g., Fourier transformations), and 3. ``Setwise'' features, which incorporate information from all spectra in the dataset (e.g., PCA).

Local features are useful when spectral features are localized to specific energy regions. They allow for easy interpretation between transformed features and energy regions, which can help identify why specific features may be important for prediction. However, they can miss global information on periodic or oscillatory trends that nonlocal features can highlight. For this reason we provide both local and nonlocal featurization techniques to complement each other for prediction. Setwise features like PCA create greater risks for extrapolation as they rely on properties of the original training dataset, but may help with understanding global properties of the dataset.

We summarize our featurization methods in Table \ref{tab:1}, and leave full descriptions of them to the supplemental information.

\begin{table}[]
\centering
\begin{tabular}{|l|l|}
\hline
\textbf{Featurization}           & \textbf{Type} \\ \hline
Interpolation                    & -             \\ \hline
Multiscale Polynomial Fitting    & Local         \\ \hline
Gaussian Peak Fitting            & Local         \\ \hline
Continuous Wavelet Transform     & Local         \\ \hline
Nonlinear expansion              & Local         \\ \hline
Fourier Transform                & Nonlocal      \\ \hline
Cumulative Distribution Function & Nonlocal      \\ \hline
Principal Component Analysis     & Setwise       \\ \hline
Feature Agglomeration            & Setwise       \\ \hline
\end{tabular}
\caption{\label{tab:1} A brief overview of available featurizations in Spectra-Scope. Explanations of the nomenclature are provided in the text.}
\end{table}

\subsubsection{Nonlinear Feature Expansion}
Because raw spectral values and featurized values may not always have a linear correlation with the response variable, nonlinear feature expansion \cite{sun_alven_2020} can help augment the input variables with a set of nonlinear transformations. For example, $x^2, \ln{x}, (\ln{x})^2, \sqrt{x},\frac{1}{\sqrt{x}}, 
\frac{1}{x}, \frac{1}{x^2}, x^{1.5}, \frac{\ln{x}}{x}$. Any feature, such as polynomial coefficients, may be used as inputs to this feature expansion. Here, we solely apply it to raw `pointwise' spectral features in one of the case studies.

\subsection{Feature Discovery and Selection}
There are several objectives in spectral featurization. The first is to find features which have a useful mathematical correlation with the output of interest, which should have a conceptual connection to the physical mechanism that generates the spectrum. Transformed features, by exploring a wider set of data representations, may be more likely to improve performance by working better with a model's functional form, such as by making the relationship between signal and response linear. Put differently, transformations may make more similar/dissimilar spectrum-property pairs closer/further in feature space which makes for easier model fitting. One model-free way to identify such relationships is linear and nonlinear correlation analysis \cite{sun_smart_2021}, using tools such as Pearson's correlation coefficient \cite{bertsekas2008introduction} and maximal correlation analysis \cite{renyi_measures_1959}. While these model-free methods exist, here, we rely on model performance to judge feature suitability. Further nonlinear transformations or combinations of features can be applied as well \cite{sun_alven_2020}, which we explore in the Case Study section below.

The second goal is interpretability. There are two primary use cases here: gaining knowledge about the system, and gaining knowledge about the model. If the user does not have an \emph{a priori} physical model for the processes that yield the response variable and/or does not know whether the response variable has a corresponding signal in the spectrum, it may be unknown which transformations are best. In this case, feature interpretability may help the user build a mental model for the underlying process and learn about the system - for example, identifying if the presence/absence of a feature in the spectrum predicts the response variable. Further, interpretability can help with model trustworthiness, as it can help the practitioner diagnose why a failure during training or inference occurs, or predict possible failures ahead of time.

\subsection{Models}
The next step is to train models to learn the relationship between the spectra (and/or their transformed forms) and a property of interest. We note as a caveat that AutoML approaches suffer from a ``winner-take-all'' approach \cite{sun_smart_2021} which means that a model that is selected based on performance may have limitations that actually make it suboptimal in practice. For example, trying every possible model increases the degrees of freedom in training and can lead to overfitting \cite{arlot_survey2009}. Here, we hope that the emphasis on sparse and interpretable models helps mitigate the risk of overfitting by making clear what features the model is predicting on. In practice, we encourage users in production to examine the models to ensure that the features of interest correlate with the underlying phenomenon.

\subsubsection{Random Forests}
Random forests have many desirable properties: (1) They can capture nonlinear relationships without intermediate transformations, making them useful for `first-pass' modeling and screening different featurizations. (2) Random forests train an ensemble of decision trees on bootstrap aggregates (bagging) of the training dataset, and the trained trees only use a random subset of the features for each feature split. This reduces correlation between trees in the ensemble and mitigates overfitting. For Spectra-Scope it means less negative impact when adding irrelevant features, assisting feature screening. (3) Finally, off-the-shelf implementations of random forests allow users to gauge the importance of different parts of the feature space in the training dataset using feature importance scores. Further interpretability via Shapley additive explanation (SHAP) analysis is possible, but is not implemented here \cite{lundberg2017unified, lundberg2018consistent}. We refer readers who are unfamiliar with random forests to Breiman, 2001 \cite{breiman_random_2001}. We use the implementation from Scikit-Learn \cite{scikit-learn}.

\subsubsection{LCEN}

The LASSO-Clip-Elastic-Net (LCEN) algorithm of Seber and Braatz, introduced and described in detail elsewhere \cite{seber_lcen_2024}, works by sequential fitting and ``clipping'' of linear regression coefficients under different regularization conditions. We briefly summarize the four main steps of the LCEN fitting procedure. First, LASSO regression \cite{Tibshirani-1996} is performed on the input features. Second, there is a clip step, where coefficients with absolute magnitude below a user-selected cutoff are set to 0. Third, elastic net regression is performed on the features that have not been clipped in step 2. Finally, a second clip step is performed on the resultant coefficients. The clip steps perform feature selection to create accurate but sparse models. The algorithm can also perform transformations on the input data to find nonlinear relationships between the data and the label. Including these transformations can be turned on or off through Spectra-Scope. Here, we use the implementation of LCEN from the Smart Process Analytics (SPA) repository (\url{https://github.com/PedroSeber/SmartProcessAnalytics}).

The clip steps, by eliminating coefficients of small magnitude, mean that any surviving nonzero coefficient is making a significant contribution to the prediction. When the features are scaled, the magnitude of the coefficients can also be used to interpret the relative contributions. Therefore, interpretability comes from examining the magnitudes of the coefficients and the user can learn more about which features are contributing the most to predicting the response variable.

\subsubsection{Fused lasso}
We include the generalized lasso or ``fused lasso'' as a built-in model to Spectra-Scope. The fused lasso is lasso regression with a penalty on the differences of adjacent coefficients \cite{tibshirani_2011}, which is well suited to spectral data. The loss function is

\begin{equation} 
\min_{\beta \in {\mathbb{R}^p}} \frac{1}{2}|y - X\beta|_2^2 + \lambda |D\beta|_1 
;\qquad D = \begin{bmatrix} -1 & 1 & 0 &\dots & 0 \\ 0 & -1 & 1  & \ddots &  \vdots \\ \vdots & \ddots & \ddots & \ddots & 0 \\ 0 & \dots & 0 & -1 & 1 \end{bmatrix} \in {\mathbb{R}}^{(p-1)\times p}. 
\end{equation}

This loss function is that of least squares regression with a penalty $\lambda$ on adjacent terms \cite{tibshirani_solution_2011}. This model has previously been used to create sparse, interpretable models for battery cycle life prediction and identify regions of interest for closer study \cite{rhyu_systematic_2025}. Minimizing the difference between adjacent coefficients lends itself to spectroscopy or time-series data because neighboring time or wavelength components are often highly correlated. This can guide partitioning of the input data into smaller sections and improve interpretability. Fused lasso models themselves may be useful, or the regions they identify can be used to build understanding of the relationships between signal and response in a dataset, as we show later in the Case Study section.

\section*{Application}
\begin{figure}
    \centering
    \includegraphics[width=0.5\linewidth]{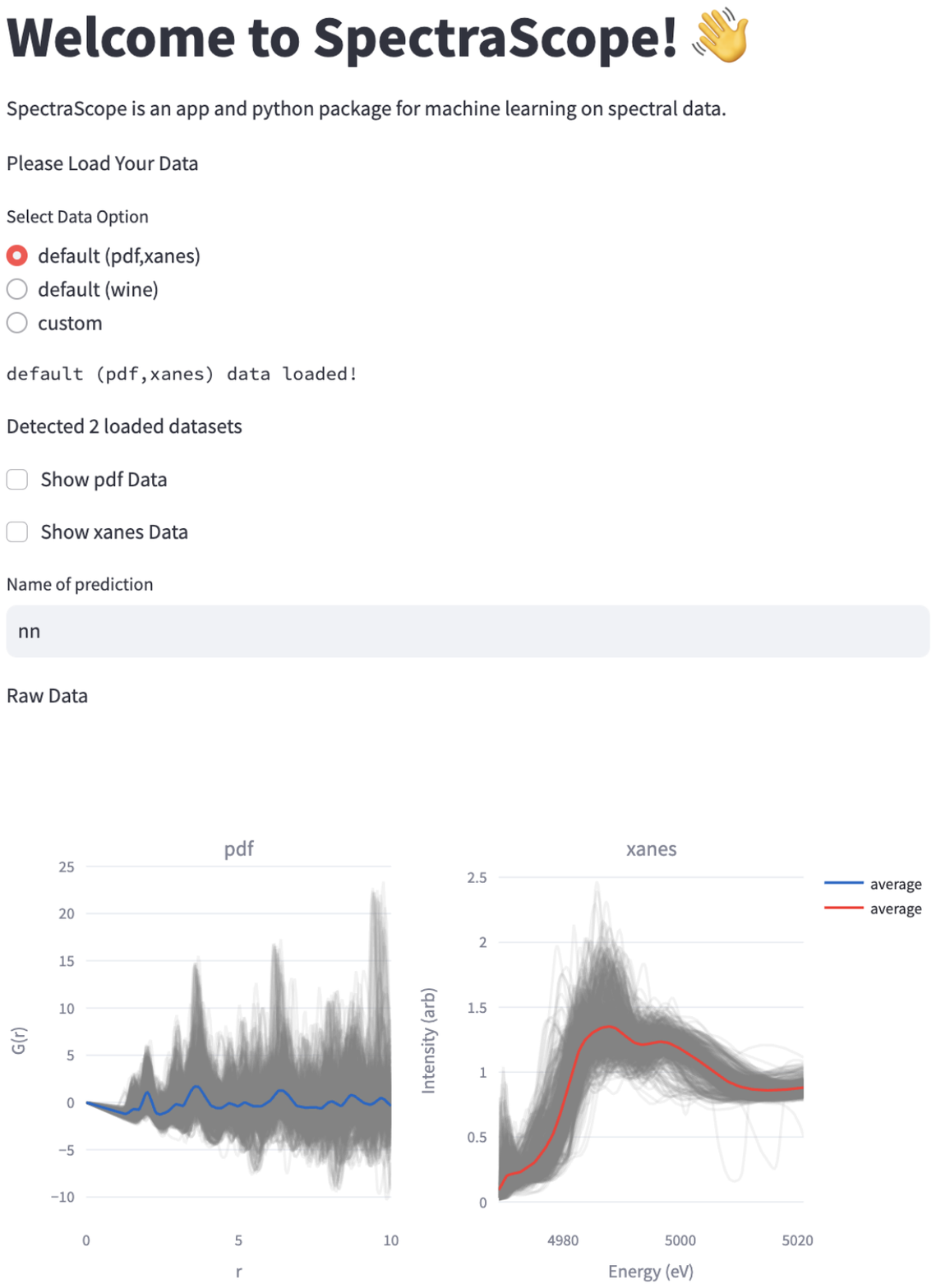}
    \caption{\textbf{Front page of Spectra-Scope application.} Multiple data types can be input and visualized on the home page. The app includes abilities to featurize data, visualize featurizations, train models using random forests or LCEN, and visualize the important or selected features by the model.}
    \label{fig:2}
\end{figure}

Spectra-Scope is available as a no-code web-based application at \href{http://spectrascope.matr.io}{spectrascope.matr.io} (See \autoref{fig:2}) implemented in Streamlit and able to be hosted locally using the code in the linked repository. After data preparation into an appropriate tabular format is done by the user, uploading data allows for feature generation and model fitting using RF and LCEN. Featurizations can be visualized in the app, as well as quick metrics about the performance of models trained on different features. Additionally, hyperparameter tuning can be done in the app. We hope that this application can be used to lower the barriers for novice users to access the tools implemented in Spectra-Scope. 

\section{Case Studies}
We benchmark Spectra-Scope on two distinct spectral datasets from the literature, recapitulating results from other studies and demonstrating how the workflow can be used to build an understanding of data.

\subsection{XANES + PDF for Transition Metal Oxides}
The first dataset consists of transition metal XAS spectra from the Materials Project \cite{jain2013commentary, mathew2018high, zheng_automated_2018}, and computed pair distribution functions (PDFs) by Na Narong \emph{et al.} \cite{na_narong_interpretable_2025}. Briefly, PDFs describe pairwise atomic distance probabilities, encoding local structural information \cite{TAKESHI201255}. XAS is an element- and orbital-specific measurement of X-ray photon absorption. Many atomic-scale properties can be inferred from XAS data, including oxidation states, density of states, coordination, and bond length in a local environment \cite{chantler_x-ray_2024}. X-ray absorption near-edge spectroscopy (XANES) measures the spectra $\approx$ 50--100 eV above the core electron energy level \cite{chantler_x-ray_2024}. Simulated and experimental XANES spectra have previously been used to predict oxidation states, bond length \cite{chen_robust_2024}, Bader charge \cite{torrisi_random_2020}, and coordination number \cite{torrisi_random_2020, zheng_automated_2018, na_narong_interpretable_2025} using machine learning models. For this case, we regress bond length from XANES and PDF data using random forests and LCEN in Spectra-Scope.

Multi-modal machine learning has proven a promising way to improve the prediction of certain materials properties using foundation models \cite{moro_multimodal_2025}. Combining features from different data sources may be used to both exploit and reveal how different techniques provide complementary information \cite{na_narong_interpretable_2025}. We highlight Spectra-Scope's ability to compare the performance of different spectral modalities in prediction, and work with multiple modalities simultaneously to perform prediction.

\subsubsection{Results}
\begin{figure}
    \centering
    \includegraphics[width=1\linewidth]{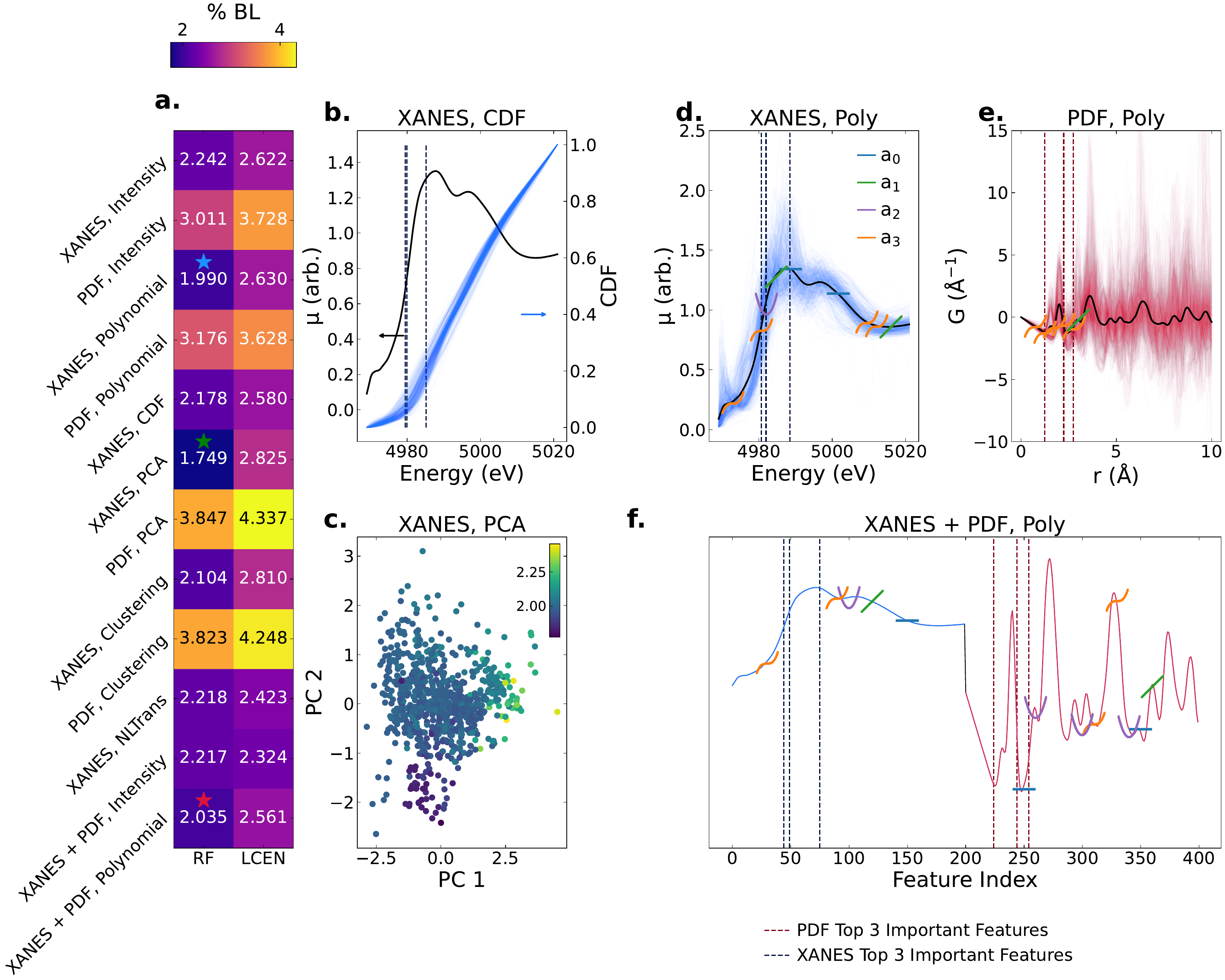}
    \caption{\textbf{Regressing mean nearest-neighbor distance from simulated
XANES spectra and PDFs of Ti-oxide structures.} (a) Summary of RMSE
for regressing bond length using LCEN and random forests for XANES, PDF,
XANES + PDF, and other transformations of the data. CDF: cumulative
distribution function. NLTrans: Nonlinear feature expansion as outlined in the
main text and the supplementary information. Clustering: Feature agglomeration clustering.
The top three features and model combinations correspond with green, blue, and
red stars, respectively. Comparison of the important features identified when
using (b) the CDF transformation of XANES spectra, (c) the first 10 principal
components of XANES spectra, and polynomial transformations of (d) XANES,
(e) PDF, and (f) XANES + PDF for regression with random forests. (b) CDF of
all XANES spectra in blue. Vertical dashed lines: top three important features
for prediction using the CDF. (c) First two principal components of the XANES
spectra colored by bond length. Color bar: bond length. (d) All XANES spectra from the dataset (blue),
average spectrum (black) and corresponding 10 most important features for
prediction (characteristic polynomial images). The top three most important
features are highlighted by the vertical dashed lines. (e) All PDFs from the dataset
(red), average spectrum (black), and positions of top 10 most important features
for prediction. The top three most important features are highlighted by the
vertical dashed lines. (f) Average XANES spectra (blue) and PDF (red) for
simultaneously using both datasets for regression. Dashed lines show the top
three most important features extracted from fitting XANES (blue) and PDF
(red) spectra separately.}
    \label{fig:Fig3}
\end{figure}

We first regress mean Ti atom nearest-neighbor distance from simulated XANES spectra of titanium oxide structures, as was performed in Na Narong \emph{et al.} and Torrisi \emph{et al.} \cite{na_narong_interpretable_2025, torrisi_random_2020}. We train both random forests and LCEN models on the featurized XANES and PDF datasets. 5-fold cross-validation is used to find the optimal model for all feature and model combinations. Full details of the model training can be found in the supplementary information. The test root-mean-square-errors (RMSEs) ranged from 0.035 to 0.088 Å, i.e., percentage errors in the range 1.74 -- 4.38 \% of the mean bond lengths. The low variance of performance across training splits in cross-validation (Supplementary Figure 1) highlights the robustness of the trained models to different splits of the data. \autoref{fig:Fig3}a provides a summary of this regression task.

Across the board, RF models perform better than the LCEN models, and some transformation schemes give slight improvement in performance over using the raw XANES intensity data. The top three performing features for RF are: the first 10 principal components of the XANES spectra, polynomial transformations of XANES spectra, and polynomial transformations of the combined XANES and PDF data. The top featurization schemes for the LCEN models are the intensity of the combined XANES and PDF data, nonlinear expansions of the XANES spectra (XANES, NLTrans), and polynomial transformations of the combined XANES and PDF data.

\autoref{fig:Fig3}(b-f) gives visualizations for a few of the different transformations used for this regression task. The cumulative distribution function (CDF) has previously been shown to give good performance on regressing bond length for simulated and experimental XANES data of Ni oxide structures \cite{chen_robust_2024}. \autoref{fig:Fig3}b shows the average XANES spectra from the dataset in black, and all of the CDFs of the spectra in blue. The vertical dashed lines give the top three most important features for prediction.

The lowest RMSE for this case comes from training a random forest on the 10 principal components of the XANES spectra. \autoref{fig:Fig3}c visualizes the first two principal components of the data, and the corresponding bond length. \autoref{fig:Fig3}d highlights the top 10 (d,e) and top 20 (f) important polynomial coefficients identified for prediction using just the XANES (d), PDF (e), or both XANES and PDF (f) datasets. The most important polynomial features selected from the XANES dataset come from features near the main edge where there is the strongest absorption. When combining the two datasets, the most important features come from regions across the datasets, rather than concentrated in particular energetic regions (in agreement with Ref. \cite{na_narong_interpretable_2025}). 

Our models perform similarly to previous studies using random forests to regress bond length of transition metal oxide structures. Na Narong \emph{et al.} \cite{na_narong_interpretable_2025} report percentage errors of 3.1--3.9\% of the mean bond lengths using the intensity data of XANES, PDFs, and combined spectra for the same structures. Torrisi \emph{et al.} \cite{torrisi_random_2020} report an $R^2$ score of 0.85 for the same regression task using polynomial transformations of XANES spectra. For the same featurization scheme, we measure an $R^2$ of 0.84. Chen \emph{et al.} \cite{chen_robust_2024} perform bond length regression on Ni-oxide structures using different featurization schemes for XANES spectra. They report test RMSE of 0.009, 0.011, and 0.020 for the raw intensity, CDF, and PCA featurizations, respectively, using random forests for regression. These compare with our values of 0.045, 0.044, and 0.035 for the same featurizations for Ti-oxide structures. We attribute differences in model performance to the different datasets used across the studies.

The magnitude of LCEN coefficients and random forest feature importances are listed for all model-feature pairs in the supplementary information. Supplementary Figure 3 compares the feature importance of a random variable with that of the other features for the trained models. For all the transformations, the majority of features generated have higher feature importance than a random variable. Supplementary Figure 4 shows a similar analysis for the absolute magnitude of coefficients selected by the LCEN model, where all selected features have coefficients larger than a random variable coefficient. This highlights the robustness of the generated features to any noise in the dataset, and the ability of the random forest and LCEN models to pick out relevant and important features for prediction.

\subsection{Raman + Vis-NIR for Wine Grapes}

Next, Spectra-Scope is applied to a publicly available experimental dataset of optical spectra from wine grapes to predict their pH and sugar content by Ebrahimi \emph{et al.} \cite{ebrahimi_advancing_2024}. The differences between this dataset and the transition metal oxide dataset highlight the flexibility of Spectra-Scope.

The dataset included Visible-Near-Infrared (Vis-NIR) and Raman spectra of two types of grape varieties, and ground truth labels of the acidity and total soluble solids (TSS) content of grapes. TSS is measured in $^\circ$Brix, where 1 $^\circ$Brix
 equals 1 gram of sucrose in 100 grams of solution. Understanding the chemical composition of grapes helps winemakers make informed decisions during growth and harvesting. Sugar content and acidity are important indicators for determining when to harvest grapes and can influence the potential alcoholic content and fermentation process of wines, respectively \cite{cramer_transcriptomic_2014}. Spectral analysis allows non-invasive, real-time measurements of grape chemistry at vineyards, reducing the need for costly off-site processing and analysis \cite{ebrahimi_advancing_2024}. Vis-NIR spectra have been applied to estimate TSS in a variety of crops \cite{jha_nondestructive_2004, Liu_2011}, and Raman spectroscopy is used to identify molecular profiles present in a sample.

\subsubsection{Results}

Ebrahimi \emph{et al.} compare the performance of different machine learning algorithms on predicting pH and TSS in two grape varieties when using either the full spectrum or the spectrum transformed using principal component analysis with 6 or 15 components \cite{ebrahimi_advancing_2024}. They found a root mean squared error (RMSE) of 5--7\% when predicting TSS of French grapes. We use this dataset to compare the performance of different transformations and models using Spectra-Scope. The \% RMSE is defined as $100 \times \text{RMSE} \times \frac{1}{y_\text{max}}$ where $y_\text{max}$ is the maximum value of the $^\circ$Brix data.

\begin{figure}
    \centering
    \includegraphics[width=0.8\linewidth]{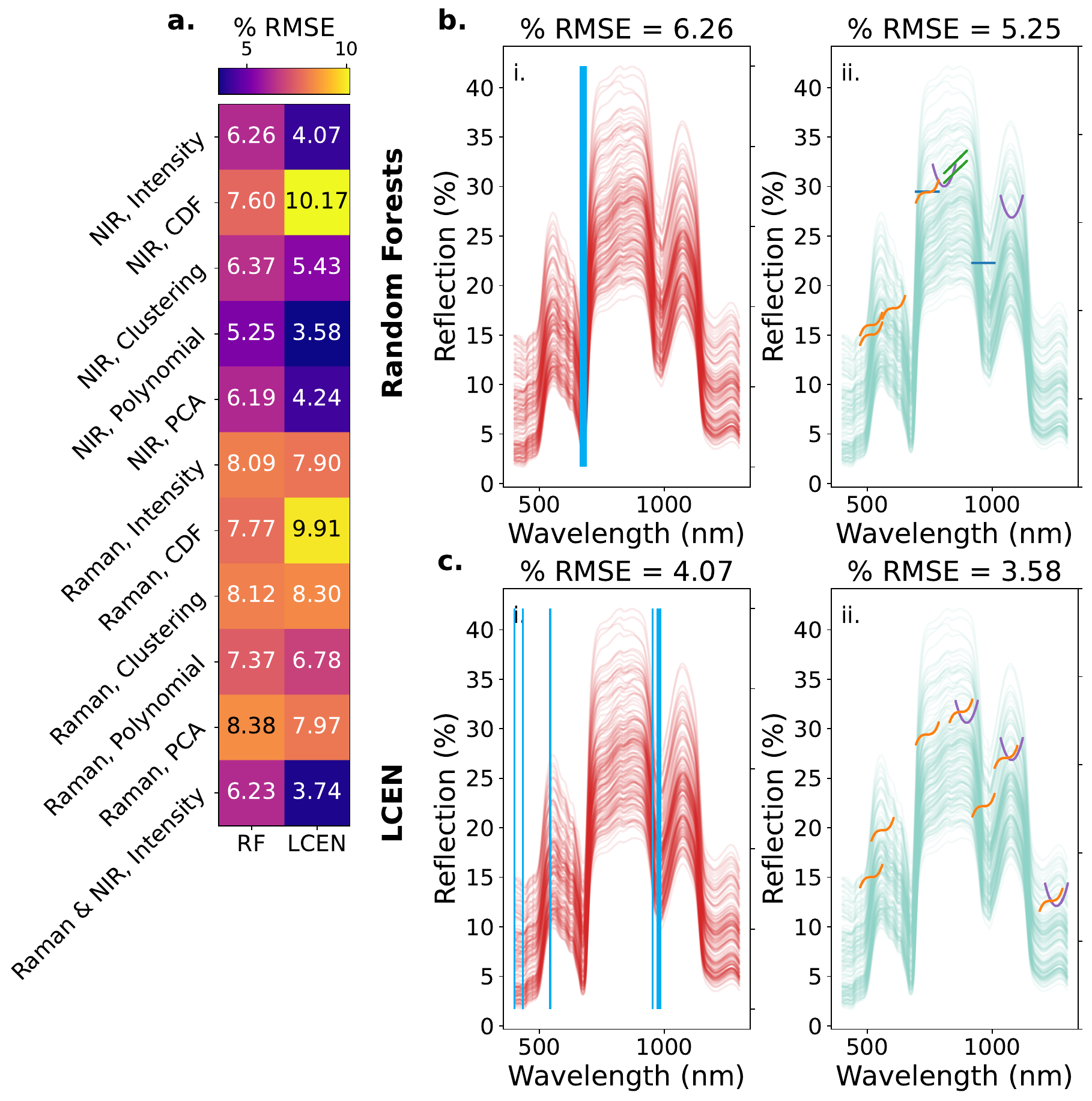}
    \caption{\textbf{Regressing grape sugar content.} (a) \% RMSE for random forests
and LCEN models built on Vis-NIR and Raman spectra transformed in different ways. (b) Top 10 most important features
for predicting TSS with the full spectrum (i) and polynomial features extracted
from the spectrum (ii) using random forests. (c) 10 highest absolute magnitude coefficients for regressing TSS with the full spectrum (i) and polynomial
features extracted from the spectrum (ii) using LCEN. Blue vertical lines: selected/important features. Characteristic polynomials: selected/important features.}
    \label{fig:Fig4}
\end{figure}

We train a series of random forest and LCEN models using Spectra-Scope to predict TSS using the Vis-NIR and Raman spectra from the dataset. We use similar preprocessing steps as Ebrahimi \emph{et al.} by centering and scaling the data, and restricting the Vis-NIR spectra to between 400 and 1300 nm, where the strongest absorption values are. The models are trained on the preprocessed intensity data or features generated by different transformations. Again, we can use Spectra-Scope to quickly visualize the performance of different models and feature transformations as given in \autoref{fig:Fig4}a. Our reported \% RMSE are similar or better than those reported by Ebrahimi \emph{et al.} \cite{ebrahimi_advancing_2024} for similar models. \autoref{fig:Fig4}a suggests that a better prediction of TSS is found when training LCEN models on the Vis-NIR absorption spectra. The top 3 performing features for the LCEN models are the polynomial transformations of the Vis-NIR spectra, combined Raman and Vis-NIR intensity data, and the preprocessed intensity of the Vis-NIR spectra.

Thus, we focus primarily on Vis-NIR spectra for our analysis. In \autoref{fig:Fig4}b and c we highlight the 10 most important features or highest magnitude coefficients using (i) interpolated data and (ii) polynomial coefficients for both the (b) Random forest and (c) LCEN models. Each featurization and model combination highlights different components of the spectra as important for prediction. In \autoref{fig:Fig4}b(i), the random forest selects wavelengths from 650--680 nm as important. In (ii), the polynomial coefficients around 550, 738, 800, 970, and 1100 nm are important. In \autoref{fig:Fig4}c(i), the LCEN model selects high magnitude coefficients around 550 and 970 nm. In (ii), polynomial coefficients around 550, 738--900, 970, 1100, and 1200 nm are selected by the model.

Many of the wavelengths selected as important by the models correspond with second or third overtones of molecular infrared vibrational frequencies that may be prevalent in grapes. The absorption peak at 970 nm likely corresponds with water, which has a near-IR absorption band near 970 nm due to overtones of the O-H stretching mode of water \cite{lin_theory_2009, golic_short-wavelength_2003}. The third overtone of this vibrational frequency is at 738 nm, which is picked up by 3 of the model-feature pairs. The small feature around 840 nm may correspond with the second overtone of the 1100 nm O-H combination band in water. The broader peak at 1200 nm may correspond with the second overtone of C-H and C-H$_2$ stretching frequencies, which would come from organic solutes like glucose and sucrose, and have been reported at 1100--1230 nm and 1215 nm, respectively \cite{golic_short-wavelength_2003}. These spectral features have previously been used for glucose monitoring \cite{golic_short-wavelength_2003}. As these vibrations are not Raman-active, this explains why the Vis-NIR spectra perform better in this analysis. 

For the interpolated and polynomial featurization schemes shown in \autoref{fig:Fig4}c, the final LCEN model selects 88 and 31 features, respectively. To contrast, random forest models predict on all the available features in a dataset, which are of size 899 and 157, respectively. This highlights the utility of regularization for feature selection, as nonzero features can be more easily judged for their importance than a feature score which is computed for all features in a random forest model.

\begin{figure}
    \centering
    \includegraphics[width=0.8\linewidth]{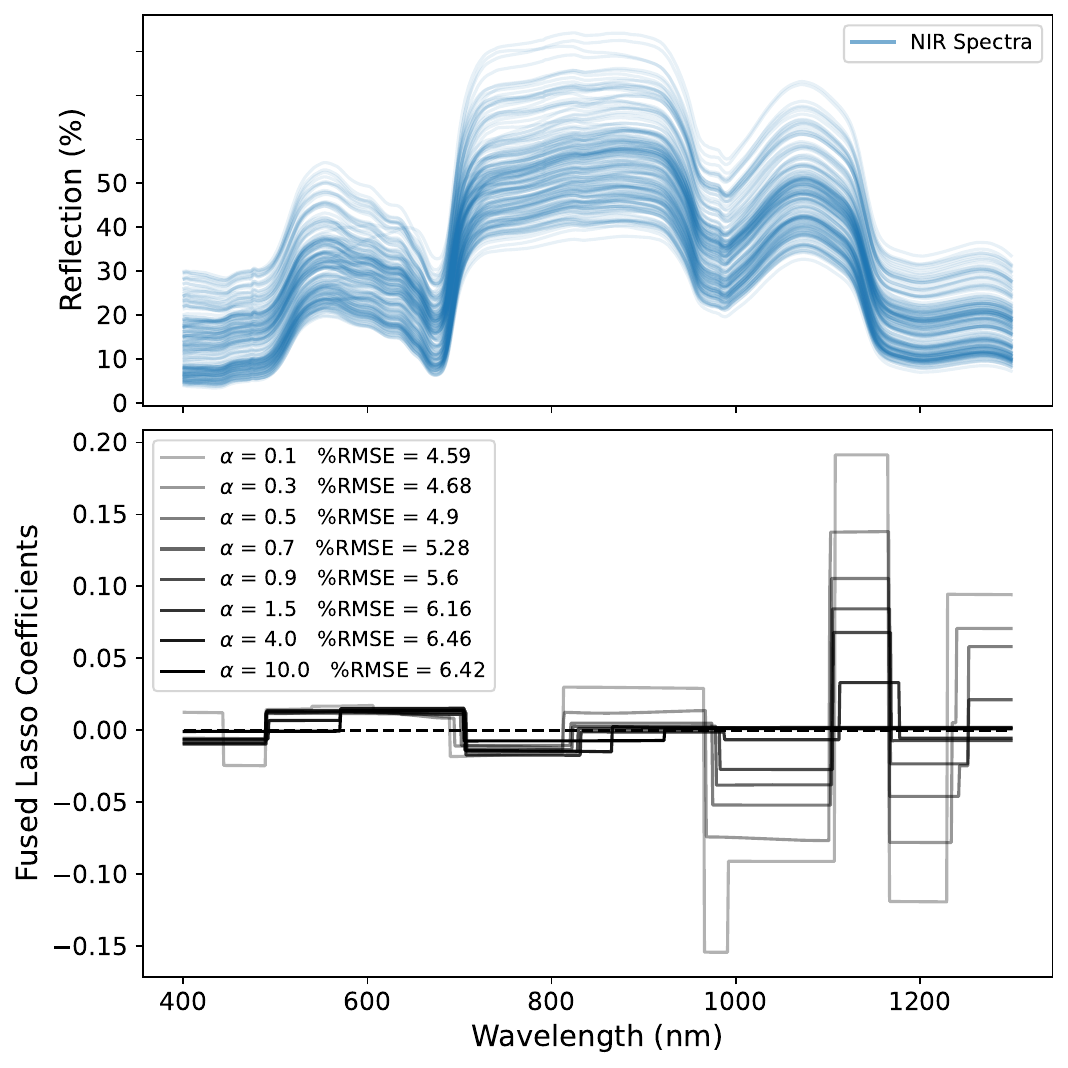}
    \caption{\textbf{Fused lasso selected features.} The top panel shows
all of the Vis-NIR spectra in the dataset. The bottom panel shows the regression
coefficients for fused lasso models with different regularization parameters $\alpha$.}
    \label{fig:Fig5}
\end{figure}

\autoref{fig:Fig5} shows the regression coefficients of a fused lasso model trained on the raw intensity Vis-NIR data. As the fused lasso penalizes differences between neighboring coefficients, the model returns regions of the spectra with the same or very similar coefficients. Positive/negative coefficient values suggest that data within these regions correlate/anticorrelate with the output. As the regularization term increases, regions that are less important for prediction will approach 0. The regions that remain ``on'' with greater regularization are around 550 nm to 700 nm and 700 nm to 850 nm. This suggests that the 738 nm overtone that was highlighted by the random forest and LCEN models in \autoref{fig:Fig4} is important for this case as well. Additionally, the coefficients change value near 840 nm, 970 nm, and 1150 nm for all regularization parameters. This highlights these wavelengths as important, as they resist the regularization penalty imposed by the fused lasso model. The fact that the most consequential (judging by the largest coefficient magnitude) regions correlate with the vibrational modes of organic solutes indicated above builds confidence in this model. Seeing model agreement across different featurization modes in certain regions can be taken as evidence that they are fitting on consistent signal. Coefficients of the best fused lasso model are in the supplementary. 

\begin{figure}
    \centering
    \includegraphics[width=0.5\linewidth]{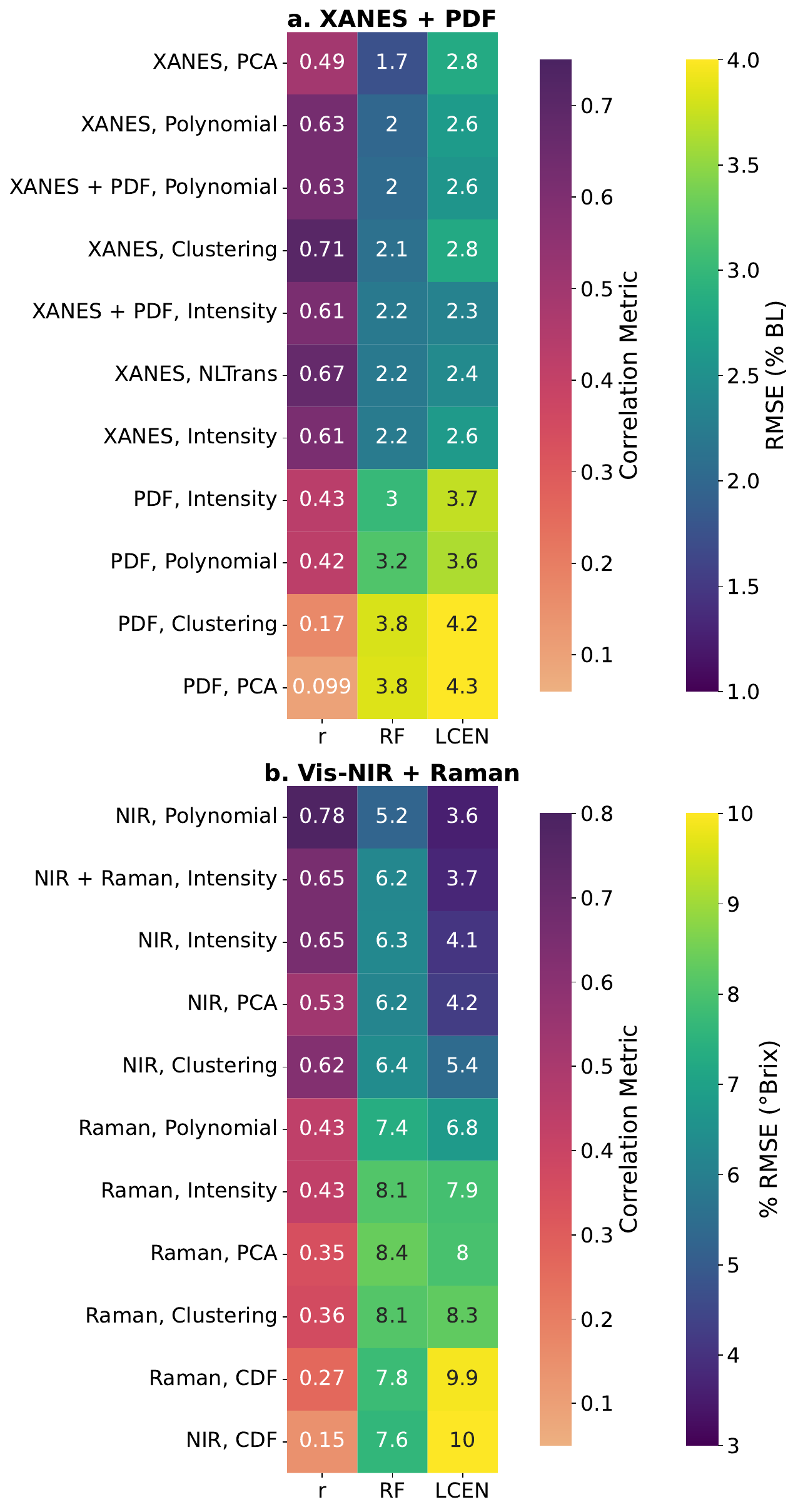}
    \caption{\textbf{Linear Correlation Assessment.} Analysis for (a) XANES + PDF
and (b) Raman + Vis-NIR dataset for predicting their respective target variable. r: Pearson’s correlation coefficient. Right: Corresponding metric for each model.}
    \label{fig:Fig6}
\end{figure}

Finally, we briefly compare feature-performance correlations across datasets in \autoref{fig:Fig6}. As a sanity check, we find that the features that perform the best with the LCEN model (and overall) have the highest Pearson's correlation coefficients in this dataset, as expected for a linear model. This suggests that LCEN models perform well due to nonlinear transformations uncovering linear correlations with the target variable. For the XANES + PDF dataset, there is less of a correlation between those features with the highest Pearson's correlation coefficients, and the lowest RMSE. This suggests that the underlying relationship is more nonlinear, and could explain why the random forests perform better than LCEN for this case study.

This highlights Spectra-Scope's ability to go beyond linear regression. Our toolkit has the power to find nonlinear spectrum-property relationships in two ways: (1) the structure of random forests allows them to capture nonlinear and non-monotonic relationships between data and labels, and (2) nonlinear transformations of features with LCEN ($x_i^2, x_i\cdot x_j, \sqrt{x_j}$) and built-in functions (polynomial coefficients, Gaussian peak fitting, etc.). Both of these handles explicitly allow the model to learn spectrum-property relationships beyond linear regression of the input variables. 

Linear correlation analysis can help determine if one's data is well suited to linear regression as a functional form. If one cannot find a linear correlation between any features generated and the target, then LCEN or fused lasso regression will not work, and random forests should be tried. Other data-driven nonlinear transformations can combine data from different features, such as neural networks. While neural networks are not studied here, if needed, a user can easily integrate a simple multilayer perceptron model from scikit-learn into the Spectra-Scope workflow, and have access to more flexible models. 

The magnitude of LCEN and fused lasso coefficients and random forest feature importances are listed for all model-feature pairs in the supplementary information. Supplementary Figure 5 compares the feature importance of a random variable with that of the other features for the trained models. Except for the worst performing feature transformations, the majority of features generated in each transformation are more important than that of a random variable, highlighting the model's ability to pick up on real signal and be robust to noise in the dataset. Supplementary Figure 6 shows a similar analysis for the absolute magnitude of coefficients selected by the LCEN model, which are primarily larger than a random variable coefficient, save for the worst performing model-feature pairs.

\section{Discussion}

These case studies highlight how the ability to compare a variety of transformations and models at once allows a Spectra-Scope user to quickly identify which model may be best for their regression or classification task. Our agreement with past literature validates that Spectra-Scope recovers physically meaningful features, supporting the toolkit's reliability for exploratory analysis. While we focus on two distinct case studies and four different types of spectra here, we imagine Spectra-Scope can be applied to any dataset where a relationship exists between spectra and materials properties. In addition to the spectrum-property relationships already mentioned and analyzed, XANES spectra can be used to predict the local structure and valence in amorphous silicon suboxide \cite{Fujikata2026}, and Raman spectroscopy could be used to detect microplastics \cite{Sunil2024}. In a very different area of research, Raman spectroscopy has been used to predict the crystal lattice misalignment, or twist-angle, of stacked MoS$_2$ \cite{Tang2023} and graphene \cite{Fernandez2022} layers. These examples highlight the broad array of tasks that Spectra-Scope could be applied to for accelerated materials research and discovery. 

Attempting different featurizations and model types and comparing which parts of spectra are used may be used for hypothesis formation and building confidence. For example, if fused lasso and polynomial featurization highlight the same regions, it may be taken as evidence that the energy ranges contain meaningful signal. When using LCEN, the subsets of data that are used by the fitting procedure should be studied as the regularization parameter changes. We see consistency in our fused lasso results, but we encourage checking this in future deployment.

There are some limitations to the AutoML approach. As mentioned earlier, using model performance alone as a ``winner-take-all'' measure of feature selection risks overfitting. Models that perform slightly worse across the same set of validation tests may work better in deployment or generalize better to out-of-distribution data if they verifiably are fitting on physically meaningful signals. Sanity checks on the feature sets employed should be performed for physical plausibility. For example, if there is expected to be periodic signal in the data, then the Fourier and wavelet transforms are more sensible choices of featurization. Depending on dataset size, it may be more desirable to perform first-pass maximal correlation analysis. We use the \texttt{ace} package in our library to perform this, but do not discuss this in the manuscript. 

Here, we only explore random forests and LCEN. Poor performance with these models doesn't mean that a data-driven approach should be disregarded. Maximum accuracy may come from more dedicated feature and model design. If more data is available, featurization techniques may not be as necessary, as one of the hallmarks of artificial neural networks is that they learn their own representations of data depending on the task at hand. They may in these cases obtain better performance and robustness. We expect featurization techniques to be most useful in a lower-data regime (e.g. on the order of hundreds to thousands of data points), where they represent a form of inductive bias introduced to the data. 

Furthermore, some use cases may be beyond the scope of Spectra-Scope. Uncertainty quantification is not formally supported. For multimodality, we demonstrate that Spectra-Scope can work with simple concatenation of feature sets. We expect that more sophisticated methods that combine information from two different datasets could be employed and integrated into Spectra-Scope in the future \cite{subramanian_xxact-nn_2025, moro_multimodal_2025}.

We highlight some failure modes for models discovered using an AutoML approach. Users should verify that adequately performing feature sets correspond with physically meaningful signals and the needs of future deployment. For example, finding correlations between the Vis-NIR regions of interest and vibrational modes that are likely to matter for $^\circ$Brix data highlights the physical relevance of the found regions. Alternatively, if regions highlighted as important are known to have low signal-to-noise ratios, then the model should be vetted for overfitting, as spurious correlations may be picked up by the model. For setwise features like PCA or hierarchical clustering, understanding the similarity between inference-time data and training data is very important.

\section*{Conclusion}

Spectra-Scope is a flexible platform for building interpretable machine learning models with spectral data. We highlight its ability to draw insights from diverse datasets and aid practitioners in better understanding their data and models. Its accessibility is enhanced by its corresponding web app, allowing it to be used by scientists with and without Python experience. While we've currently focused on spectral data, Spectra-Scope can take any array-like data as input, allowing it to be used with other characterization techniques, and applied to problems beyond materials science. 

We hope that a focus on sparsity and interpretability will make for easier deployment of models in production in contexts ranging from automated materials discovery and high-throughput experimentation to industrial manufacturing. For example, models which operate on only a few features may be easier to verify, troubleshoot, and deploy to operate without supervision. For future users, Spectra-Scope is an open-source Python package that is available at \href{http://github.com/TRI-AMDD/spectrascope}{github.com/TRI-AMDD/spectrascope}, with a web application hosted at \href{http://spectrascope.matr.io}{spectrascope.matr.io}.
\bibliography{ss}

\section*{Acknowledgements}

The authors thank Pedro Seber for advice on the implementation of LCEN, Richard Braatz, Tina Na Narong, and Simon Billinge for helpful discussions, and Linda Hung for helpful feedback on the manuscript.

\section*{AI Usage Disclosure}
Grammar, concision, and typo checks were performed using Claude Opus 4.5. Generated suggestions were considered but not necessarily applied, and all changes were directly mediated by authors after review by hand. All authors have reviewed and approved all information demonstrated in this work.

\section*{Author contributions statement}

S.B.T. conceived the project. A.C.J. developed the library, trained the models, and interpreted the results with the guidance of S.B.T. and W.Y. C.F. developed the application with help from L.S. A.C.J. led the drafting of the manuscript, with editing and revisions from S.B.T. and feedback from W.Y.
All authors reviewed the manuscript. 

\section*{Competing Interests}
The authors have no competing interests to declare.

\section*{Data Availability Statement}
Spectra-Scope is an open-source Python package made available at \href{http://github.com/TRI-AMDD/spectrascope}{github.com/TRI-AMDD/spectrascope}, with a web application hosted at \href{http://spectrascope.matr.io}{spectrascope.matr.io}. 

\section*{Funding Statement}
This work was funded by Toyota Research Institute. 
\end{document}